\documentclass[amsmath,twocolumn,aps,prl]{revtex4}
\usepackage{bm}
\usepackage{graphics}
\begin{document}
\title{Reply to Comment: S. Das Sarma et al., arXiv:0708.3239}
\author{E.G.~Mishchenko}
\affiliation{Department of Physics, University of Utah, Salt Lake
City, Utah 84112, USA}
\begin{abstract}

The authors of the Comment ascribe us claims never made while
moderating their own previous unsubstantiated statements.

\end{abstract}


\maketitle

Ref. [1] states:\\
i) "{\it recent Letter [2] by Mishchenko is both incorrect and
misleading}".\\
ii) "{\it we emphasize here that for extrinsic graphene (i.e. gated
or doped graphene with a free carrier density-induced chemical
potential or Fermi level $E_F$ in the conduction or valence band
away from the Dirac point) the RPA is an excellent and controlled
approximation}".\\
iii) "{\it the sweeping statement made in Ref. [2] about the lack of
validity of RPA in calculating the graphene self-energy, even in the
weak-coupling $r_s \ll 1$ regime, is thus incorrect for extrinsic
graphene and trivial for intrinsic graphene}". 
\\

The criticism of Ref.~[1] is misaddressed. Though the above are
strong assertions, they obviously have no relation to our paper [2]
which addresses undoped ('intrinsic') graphene, the fact repeatedly
stated throughout the paper (including the abstract). The paper [2]
contains no statements whatsoever about doped ('extrinsic')
graphene. In particular, it finds and unambiguously states that
($g\equiv r_s=e^2/\hbar v$) "{\it the neglect of non-RPA corrections
to the electron self-energy becomes an uncontrollable approximation
in the higher orders in $g$ for an undoped graphene}".

Ref.~[2] points out that "{\it our findings, thus, do not support
the conjecture of Refs.~[3-4] that RPA is exact approximation in the
limit $g \ll 1$}". Indeed, our calculations in Ref.~[2] demonstrate
that the sweeping statement made in Ref. [3] about the validity of
RPA in the limit $g \ll 1$ for both extrinsic and intrinsic graphene
is unfounded. It is worth quoting the claim made in Ref.~[3]
verbatim:

"{\it In conclusion, we have presented a calculation, formally exact
in the $r_s \ll 1$ limit, for the renormalized Fermi liquid
parameters for both extrinsic and intrinsic graphene}".

The quote unambiguously shows that the 'trivial' fact that RPA lacks
validity in intrinsic graphene was not recognized by the authors of
Ref.~[3] at publication. Neither was it recognized in the later
publication \cite{4} which repeats indiscriminately that RPA "{\it
should be an excellent approximation for graphene since graphene is
inherently a weak-coupling (or "high-density" in parabolic-band
systems) 2D system [3]}".

In fact, parameter $g$ (being the same in both the intrinsic and
extrinsic graphene) alone is insufficient to distinguish whether RPA
is valid or not for finite doping. A proper condition should relate
the relevant energy scale (e.g.~the arguments of the electron
self-energy) to the Fermi energy (degree of doping). The references
\cite{3,4} do not offer such a criterion. Ref.~\cite{1} moderates
previous sweeping claims, and rightly so, to the extent that "{\it
RPA is a perfectly meaningful approximation for extrinsic (i.e.
doped or gated) graphene}", still Ref. [1] is quiet  on what
specific numerical condition should distinguish extrinsic (where RPA
is valid) from intrinsic  case (where it fails 'trivially').

The calculations presented in the Comment [1] for finite $E_F$
(extrinsic graphene) arrive at the well-known fact \cite{Mahan} that
RPA-loops are singular at small momenta and confirm our conclusion
(yet deemed 'incorrect and misleading' in the first line of the
Comment) that for $E_F=0$ (intrinsic graphene) non-RPA terms are of
the order of RPA terms \cite{2}.

However, the authors of Ref.~[1] were yet again unable to elucidate
what condition the doping level $E_F$ should satisfy for the RPA to
be valid.


\begin{thebibliography}{50}
\bibitem{1} S. Das Sarma, Ben Yu-Kuang Hu, E. H. Hwang, Wang-Kong Tse,
arXiv:0708.3239.

\bibitem{2} E.G. Mishchenko, Phys. Rev. Lett. {\bf 98}, 216801 (2007).

\bibitem{3} S. Das Sarma, E.H. Hwang, and Wang-Kong Tse, Phys. Rev. B {\bf 75},
121406(R) (2007).

\bibitem{4} E.H. Hwang, B.Yu-K. Hu, and S. Das Sarma,
cond-mat/0612345.

\bibitem{Mahan} G. Mahan, {\it Many-particle physics} (Kluwer, New
York, 1981).

\end{thebibliography}
\end{document}